\documentclass{ws-jcsc}
\usepackage{amsmath,amssymb,amsfonts}
\usepackage{algorithm}
\usepackage{algorithmicx}
\usepackage{bm}
\usepackage{booktabs}
\usepackage{multicol}
\usepackage{graphicx}
\usepackage{textcomp}
\usepackage{float}
\usepackage{stfloats}
\usepackage{tabularx}

\begin{document}

\markboth{Jintao Yu, Bing Xiao, Yuzhu Cui.}{Robustness of double-layer group-dependent combat network with cascading failure}

%
\catchline{}{}{}{}{}
%

\title{Robustness of double-layer group-dependent combat network with cascading failure\footnote{This work is partially supported by the National Foundation of Nature Science under Grant No.~61502522.}}

\author{Jintao Yu\footnote{Corresponding author.}}

\address{Air Force Early Warning Academy\\
Wuhan, 430019, China\\
$^\dag$haleine@uste.edu}

\author{Bing Xiao}

\address{Air Force Early Warning Academy\\
Wuhan, 430019, China\\
xb\_sky@126.com}

\author{Yuzhu Cui}

\address{Tsung-Dao Lee Institute, Shanghai Jiao Tong University\\
Shanghai, 210022, China\\
yuzhu\_cui77@163.com}
\maketitle

\begin{history}
\received{(Day Month Year)}
\revised{(Day Month Year)}
\accepted{(Day Month Year)}
\end{history}

\begin{abstract}
The networked combat system-of-system (CSOS) is the trend of combat development with the innovation of technology. The achievement of combat effectiveness requires CSOS to have a good ability to deal with external interference. Here we report a modeling method of CSOS from the perspective of complex networks and explore the robustness of the combat network based on this. Firstly, a more realistic double-layer heterogeneous dependent combat network model is established. Then, the conditional group dependency situation is considered to design failure rules for dependent failure, and the coupling relation between the double-layer subnets is analyzed for cascading failure. Based on this, the initial load and capacity of the node are defined, respectively, as well as the load redistribution strategy and the status judgment rules for the cascading failure model. Simulation experiments are carried out by changing the attack modes and different parameters, and the results show that the robustness of the combat network can be effectively improved by improving the tolerance limit of one-way dependency of the functional net, the node capacity of the functional subnet and the tolerance of the overload state. The conclusions of this paper can provide a useful reference for network structure optimization and network security protection in the military field.
\end{abstract}


\section{Introduction}
Informatization warfares are systematic fighting based on information systems, and information has increasingly become the dominant factor in combat. In actual operations, one or several nodes in the combat network may fail due to combat attacks or random failures. The service processing load of these failed nodes will be transmitted through the information flow and cause secondary failures in other nodes. A larger-scale chain effect caused by the load reallocation will eventually lead to partial failure or even complete collapse of the network. In the current situation, the components and structural relations of the modern combat network are increasingly complex, and information interactions are becoming increasingly frequent. Therefore, studying the characteristics and laws of cascading failure of the combat network through reasonable models is of great importance, which is also beneficial to reducing the risk of cascading failure and making the combat network more robust.

In recent years, scholars have proposed many different models to study network cascading failure. For example,
Kinney et al.~\cite{Kinny2005} proposed a cascading failure model based on efficiency analysis for the power grid, considering the situation that the node does not disappear after failure. Wang et al. \cite{WangZF2012} proposed a stochastic Markov model, which is able to capture the progression of cascading failures by a flow redistribution model. 
One of the classical cascading failure models is the `Capacity and Load' model presented by Motter and Lai~\cite{MLmodel2002}. From the way the load is redistributed in this model, Duan et al.~\cite{DuanDL2013} proposed a complex network cascading failure model with an adjustable load redistribution range considering the load redistribution heterogeneity, and analyzed the cascading failure conditions on a scale-free network. Yin et al.~\cite{YinRR2014} also proposed the cascading failure model based on the characteristics of the changeable load and the fixed capacity of the node in the scale-free network.

The above modeling and load redistribution methods are all based on single-layer networks. As for multi-layer networks,
Yuan~\cite{YuanM2014} introduced a cascading failure model for the complex network with a hierarchical structure. The redistribution method of the model takes the hierarchy and heterogeneity into consideration, of which the network tends to redistribute extra load to intact nodes of the same or higher hierarchies. Ben-Haim\cite{Yakov2019} studied the hierarchical network with unity of command and discussed how to design the network to manage cascading failures adequately.
Multi-layer networks are usually characterized by dependencies. Inspired by this, Buldyrev et al.~\cite{Buldyrev2010} firstly proposed the cascading failure model of interdependent networks. Adnan et al.~\cite{Adnan2021} reported a
hybrid probabilistic modeling method to balance load flow and an assessment algorithm to describe the transient stability in multiple interdependent power grid cascading failures. Zakariya~\cite{Zakariya2023} discussed the interdependent power networks' cascading failure models in terms of features, limitations, and computational speed. To mitigate cascading failures, Smolyak et al.~\cite{Smolyak2020} proposed an intuitive and simple method of protecting the critical nodes, and similar approaches were proposed by Wang et al.~\cite{WangB2020} and Shen et al.~\cite{ShenY2020}. These robustness and cascading failure studies of multi-layer networks take little account of the functional heterogeneity of the nodes.

As for the research on combat networks, Guo et al.~\cite{GuoXC2017} constructed a cascading failure model in command and control networks and analyzed the influence of load parameters, capacity parameters, and evolution step size on cascading failure invulnerability. Zhang et al.~\cite{ZhangQ2021} investigated the dynamic load redistribution strategy based on the node's local load rate with respect to the cascading failure transmission of the equipment support network. These works are more focused on cascading failures in single-layer combat networks, similar to complex networks in general, there are also more complex relations among different types of combat networks and the coupled networks are well worth investigating. Hence, Yang et al.~\cite{YangYH2017} firstly explored the cascading failure characteristics and laws of information flowing in combat systems by abstracting network structure in a hierarchical way. Wang et al.~\cite{WangZ2021} proposed the model of the military information system of system based on function dependency and analyzed the center of gravity. These works only consider the robustness of the combat network under cascading failures in terms of topology structure, but rarely address the functionality of the combat network.

According to the above analysis, although existing research has deepened the theoretical understanding of cascading failures in combat networks, there are still some deficiencies that need to be addressed: (1)~The common research mainly focused on the single-layer network, and there are few kinds of research on the cascading failure model of the combat network with hierarchical structure and coupling characteristics. (2)~The research on hierarchical structure modeling and dependent failure of combat networks needs to be further explored as well as the heterogeneity of combat networks. (3)~The method of load redistribution needs to consider the actual situation, as the load is not simply distributed equally among all neighboring nodes. To address these problems, we are going to establish a heterogeneous dependent network model with a double-layer structure according to the actual situation of the combat network, design the rules of dependent failure and overload cascading failure, and discuss the robustness of the combat network through simulation experiments.

This paper is organized as follows. In Section~\ref{sec:2}, the double-layer combat network model with group-dependent characteristics is established based on real situations at first. In Section~\ref{sec:3}, the failure model combing conditional group-dependent failure model and cascading failure model is proposed. Attack modes are also classified according to different objects and intentions. In Section~\ref{sec:4}, some experiments based on the simulated combat networks are carried out to analyze the influence of different parameters and attack modes on system robustness. The conclusions and the future work follow in Section~\ref{sec:5}.

\section{Double-layer Group-dependent Combat Network}
\label{sec:2}

To be as real as possible, the combat network is modeled from two aspects. One is the physical net in terms of physical combat equipments, and the other is the functional net in terms of logic.
For the combat physical net, each piece of equipment is deployed in a dispersed manner in physical locations and undertakes the task of energy and material transmission dominated by information flow. Therefore, combat equipments are the basis for communication and information interaction.
An information grid network that achieves high data sharing, efficient information interaction, dynamic port access, and flexible combination requirements is the result of the deep fusion of the above equipments.
Considering that the main function of a physical net is communication, the definition of it is given as follows:

\begin{definition}
    Suppose $G_{\rm W}=(V, W_a, E, W_b)$ is the physical net of combat system-of-system~(CSOS), in which the node set is $V=\{v_1, v_2, \cdots, v_{N_{\rm W}} \}$, $v_i$ represents the combat equipment with communication function in the physical net, and $N_{\rm W}$ is the number of nodes. ${W_a} = \{ {w_i}|{v_i}\} $ represents the attributes of the node, including its initial load and bearable capacity. The set of edges is $E = \{ {e_1},{e_2}, \cdots ,{e_{{M_{\rm{W}}}}}\} $, where ${e_i} = \{ {e_{jk}}|{v_j} \times {v_k}\} $ represents the communication relations among different nodes, and ${M_{\rm{W}}}$ is the number of edges. The existence of edge is mainly determined by infrastructure deployment, affiliation, and mission requirements. It is relatively fixed in general and will be flexible and changeable when carrying out missions. ${W_b} = \{ {w_{jk}}|{e_{jk}}\} $ is the attribute of the communication edge, such as bandwidth and delay.
\end{definition}
For simplicity, it is assumed that the physical net is fixed and undirected and ${W_b} = 1$, so the initial perturbation of the network topology and the performance on edges will not be considered. Denote the node of the physical net as node C.

As for functional net, the combat units and relations among them in CSOS are usually abstracted into the nodes and edges of the complex network, namely, `O, P, D, A' nodes and the corresponding edges~\cite{lanyushi2013}. The `O, P, D, A' nodes represent the intelligence obtaining unit, intelligence processing unit, decision and command unit, and attack or damage unit, respectively, and more details on structural abstraction can be found in~\cite{lanyushi2013}. There are complex interaction relations among different units, forming a network structure of heterogeneous components, multi-point interaction, multi-domain fusion, and dynamic evolution, so the heterogeneous information network is used for modeling. The functional net of CSOS with heterogeneous information is defined as follows:

\begin{definition}
    Suppose ${G_{\rm{G}}} = (V, {W_a}, E, {W_b};\varphi, \psi; {V_{\rm{G}}},{E_{\rm{G}}})$ is the functional net of CSOS, where the node set is $V=\{v_1, v_2, \cdots, v_{N_{\rm G}} \}$, and $N_{\rm G}$ is the number of nodes. ${W_a} = \{ {w_i}|{v_i}\} $ represents the service attributes of the functional node, including the service category, processing load and affordable capacity undertaken by the node. $E = \{ {e_1},{e_2}, \cdots ,{e_{{M_{\rm{G}}}}}\} $ is the edge set, in which ${M_{\rm{G}}}$ is the number of edges. ${W_b} = \{ {w_{jk}}|{e_{jk}}\} $ is an attribute of the edge, indicating the interaction strength of service information. Both nodes and edges have characteristics of type. The type set of nodes is ${V_{\rm{G}}}$, and there is a mapping function $\varphi :V \to {V_{\rm{G}}}$ that satisfies $\varphi ({v_i}) \in {V_{\rm{G}}}$, while type set of edges is ${E_{\rm{G}}}$ with mapping function $\psi :E \to {E_{\rm{G}}}$ satisfying $\psi ({e_i}) \in {E_{\rm{G}}}$. If $|{V_{\rm{G}}}| > 1$ or $|{E_{\rm{G}}}| > 1$, then the functional net of CSOS is called heterogeneous combat functional net~(HCFN).
\end{definition}

Similarly, for the sake of simplicity, this paper assumes that the service interaction strength of the edge of the HCFN is within the acceptable range, so the attribute characteristics of the edge will not be considered. In addition, the type set of nodes is ${V_{\rm{G}}}=\{${O, P, D, A}$\}$, and the type set of directed edge is ${E_G} =\{$O -- O, O -- P, P -- P, P -- D, D -- D, D -- A$\}$ .

It is notable that the functional net is unidirectionally dependent on the physical net to achieve combat utility. Although the obtained and processed intelligence information and command orders are all circulated in the functional net, the information exchanges among the nodes are all based on communication units in practice. 
Apart from the physical net and the functional net, there is another important network called the dependent net. To build the dependent relation between CSOS, the following two assumptions need to be specified:

(1) The node in CSOS only have a single function, that is, node C of the physical net can only perform information transmission, node O of the functional net can only perform information acquisition, node P can only perform information processing, node D can only perform commands and decisions, and node A can only perform combat attack.

(2) Although the physical net is abstracted from the actual equipment, it is assumed that the constraints of space and time are neglected when it undertakes communication missions. Therefore, the connection conditions will not be taken into consideration for how the functional net depends on the physical net.

On the basis of the above assumptions,  we can construct a double-layer heterogeneous dependent combat network~(DHDCN) of CSOS when the functional net is relying on the physical net, which is defined as follows:

\begin{definition}\label{def:twolayer}
  Assuming that the physical net of CSOS is $G_{\rm W}$, the corresponding functional net of is $G_{\rm G}$. The coupling relation between two networks is ${E_{\rm{D}}} = \{ {E_{{\rm{GW}}}},{E_{{\rm{WG}}}}\}$, which means that the functional net relies on the physical net. When the dependency node $v_{\rm{G}}^i \to v_{\rm{W}}^j$, let ${E_{\rm{D}}}(v_{\rm{G}}^i,v_{\rm{W}}^j) = 1$\,(for the convenience of calculation, ${E_{\rm{D}}}(v_{\rm{W}}^j,v_{\rm{G}}^i)$is also equal to $1$). The dependency nodes and edges form a dependent net as $G_{\rm D}$. All networks above together constitute the DHDCN of CSOS, which is denoted as
    \begin{equation}
      G = ({G_{\rm{G}}},{G_{\rm{D}}}{\rm{,}}{G_{\rm{W}}}).
    \end{equation}
  The adjacency matrix of DHDCN is expressed as
    \begin{equation}
      \bm{S} = \left[ {\begin{array}{*{20}{c}}
          {{\bm{S}_{\rm{G}}}}&{{\bm{S}_{{\rm{GW}}}}}\\
          {{\bm{S}_{{\rm{WG}}}}}&{{\bm{S}_{\rm{W}}}}
          \end{array}} \right],
    \end{equation}
    where $\bm{S}_{\rm G}$ represents the adjacency matrix of the functional net, correspondingly $\bm{S}_{\rm{W}}$ and $\bm{S}_{\rm{WG}}(\bm{S}_{\rm{GW}})$ are the adjacency matrices of the physical net and interdependent net.
\end{definition}
It is worth noting that the above-mentioned dependencies between heterogeneous coupling networks can be the type of one-to-one, one-to-many, and many-to-one~\cite{YuMing2020}. A form of one-to-many dependency is called group dependency, therefore a DHDCN with a one-to-many dependency is referred to as a double-layer group-dependent combat network~(DGCN).

\begin{example}
  According to the network model described in Definition~3
  , the network structure of the DGCN for CSOS is given for a certain combat scenario, as shown in Figure~\ref{fig:hete_yilai}. In the figure, the functional net is composed of nodes and edges corresponding to four combat units including O, P, D, and A, and the physical net is composed of nodes corresponding to different communication support units. Based on the dependent net, the functional net and the physical net form asymmetric coupling relations.
  \begin{figure}[!htbp]
    \centering
    \includegraphics[width=0.9\linewidth]{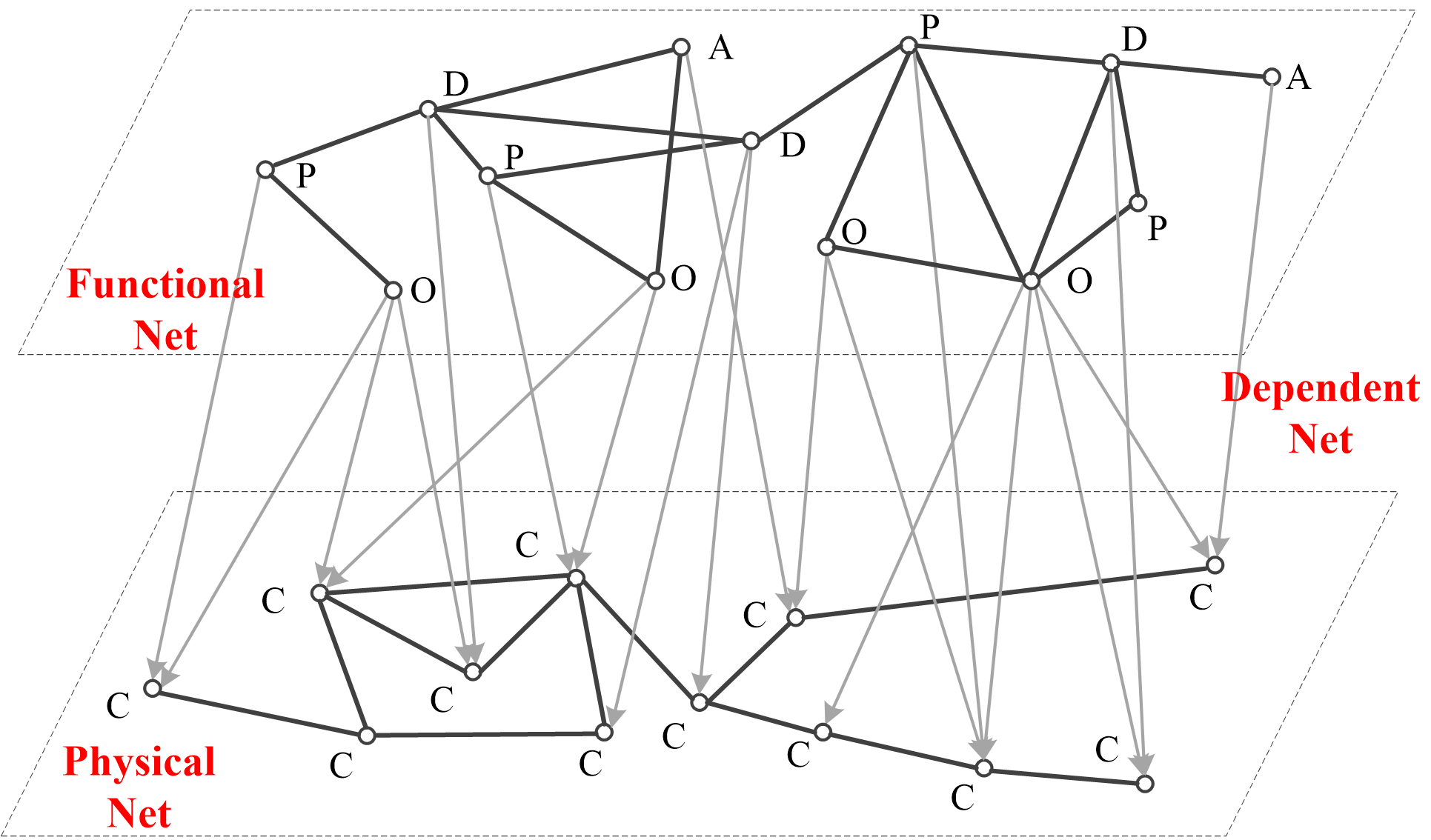}
    \caption{Structure diagram of a DGCN of CSOS.}
    \label{fig:hete_yilai}
  \end{figure}
\end{example}

\begin{figure}[!htbp]
  \centering
  \includegraphics[width=0.8\linewidth]{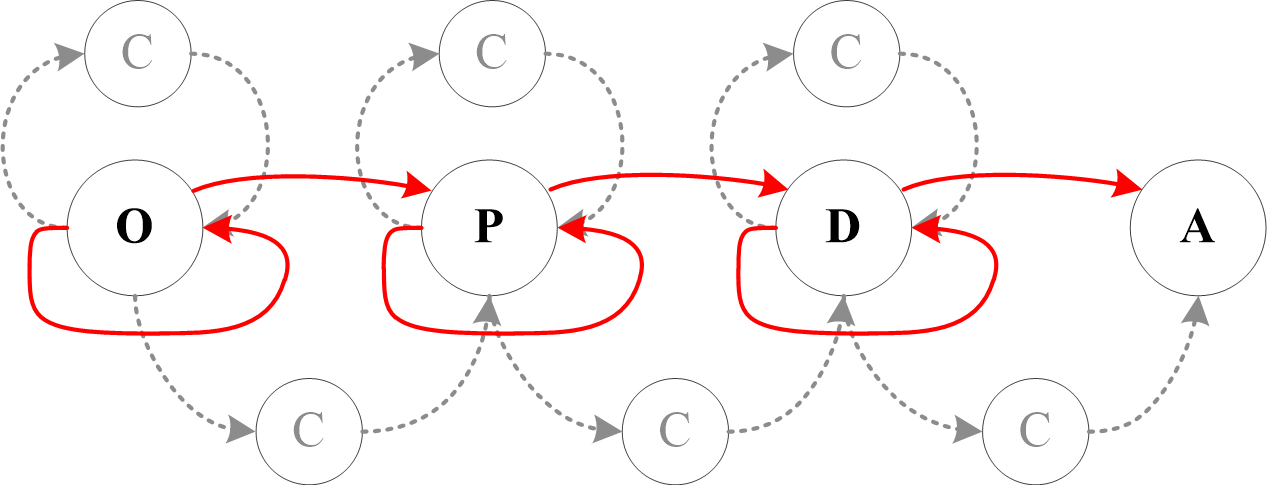}
  \caption{Information flow model based on communication nodes.}
  \label{fig:flow}
\end{figure}

The widely studied combat cycle model takes the loop composed of the target node T, search node S, decision node D, and influence node I as a measure of combat capability~\cite{TanYJ2012}. Inspired by this work, the functional net of CSOS in this paper also presents the flow of information when relying on the physical net's communication nodes. The information flow model based on communication nodes is shown in Figure~\ref{fig:flow}. As we can see from the figure, the red solid lines represent the entire information flow of combat links, and the gray dotted lines indicate the information transmission through the physical net. According to the dependency rule, the information flow of the most typical combat link `O~$\to$~P~$\to$~D~$\to$~A' will be transformed into `O~$\to$~C~$\to$~P~$\to$~C~$\to$~D~$\to$~C~$\to$~A', and other types of combat link can also be obtained similarly. Once the communication node in Figure~\ref{fig:flow} is broken, the combat link will be destroyed and CSOS will lose its combat capability. Therefore, the coupling relation between the functional subnet and physical subnet is tighter and more important in the DGCN of CSOS.

\section{Cascading Failure Model for DGCN}
\label{sec:3}

Most common interdependence network models are based on one-to-one dependent pattern, which has no feedback characteristics and is convenient for analysis~\cite{Buldyrev2010}. Quite a lot of studies on the robustness of interdependent networks are also based on this pattern~\cite{KongL2017,Dimuro2016,YuanX2017,Rocca2018}, but such strict dependencies often do not exist in practice. The situation of one-to-many dependency is relatively common, namely, the aforementioned group dependency~\cite{Wanghui2015}. If any one of the depended nodes~(physical net nodes) fails, the dependent nodes~(functional net nodes) fail according to the traditional analysis method. Therefore, group dependency greatly affects the robustness of the interdependent network. As long as a small number of nodes are attacked, the entire interdependent network may collapse, but the network in real life is not so fragile\cite{Wanghui2018,ZhangM2021}. In addition to wired communication, there are many other communication methods for the physical net of DGCN, such as short-wave communication, ultra-short-wave communication, and satellite communication. Therefore, when the functional net depends on the physical net, as long as a certain percentage of the depended nodes are still working well, the dependent functional nodes will not fail. To build the cascading failure model of DGCN, the failure analysis are as follows:

\subsection{Asymmetric dependent failure}
\begin{figure*}[!htbp]
  \centering
  \includegraphics[width=0.85\textwidth]{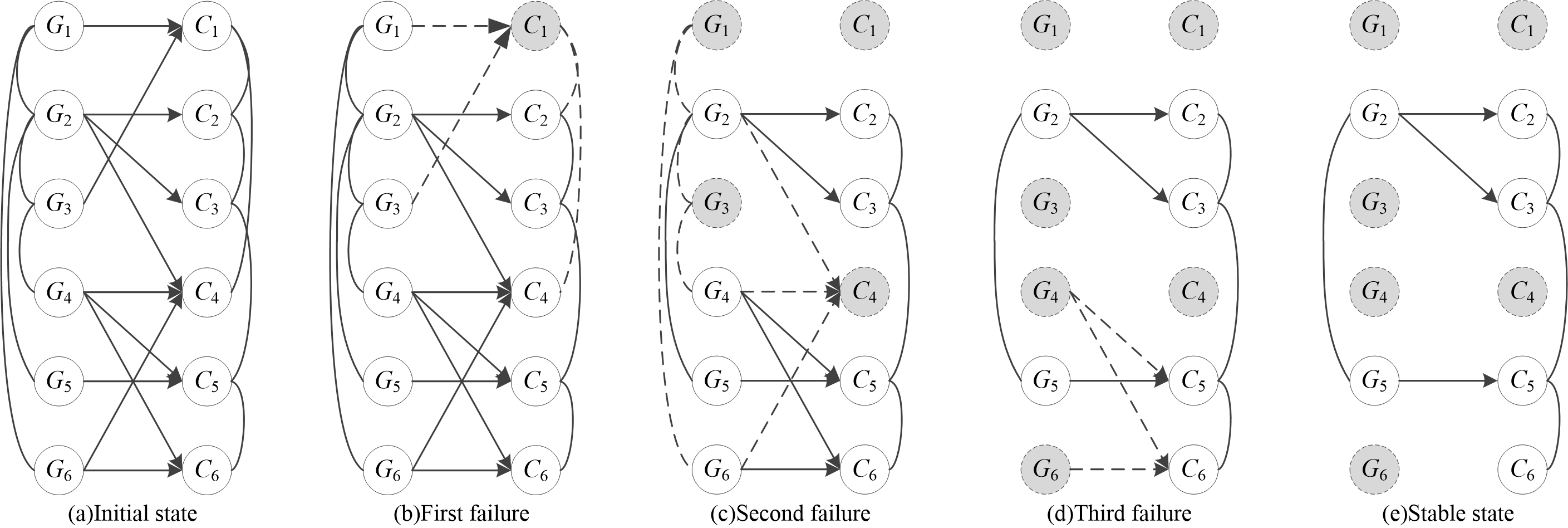}
  \caption{Asymmetric dependent failure process.}
  \label{fig:depend}
\end{figure*}

An asymmetric dependent network is a one-way dependent network. According to the classical dependent failure model~\cite{Motter2004}, the asymmetric dependent failure rules for complex networks are given: For a node that depends on the other subnet, when all its dependent nodes fail or it is not in the maximal component, the node fails; for a node of the relied subnet, when it is not in the maximal component, the node fails.
Figure~\ref{fig:depend} shows a schematic diagram of the asymmetric dependent failure process of a DGCN of CSOS.
The relied physical net node $C_1$ is initially attacked and then fails. All edges connected to $C_1$ are removed. In the second failure, the functional net nodes $G_1$ and $G_3$ that depend on $C_1$ fail due to the previous failure. Node $C_4$ also fails because it is not in the maximal component. In the third failure, nodes $G_4$ and $G_6$ fail because of the disconnection with $G_1$ and $G_3$ and they are no longer in the maximal component of the functional network. Node $G_2$ will stay normal although its connection with node $C_4$ is interrupted. Since there are no more nodes that meet the failure rules, the failure process stops and a stable state is reached.

\subsection{Conditional group-dependent failure}
\begin{figure}[!htbp]
  \centering
  \includegraphics[width=0.85\linewidth]{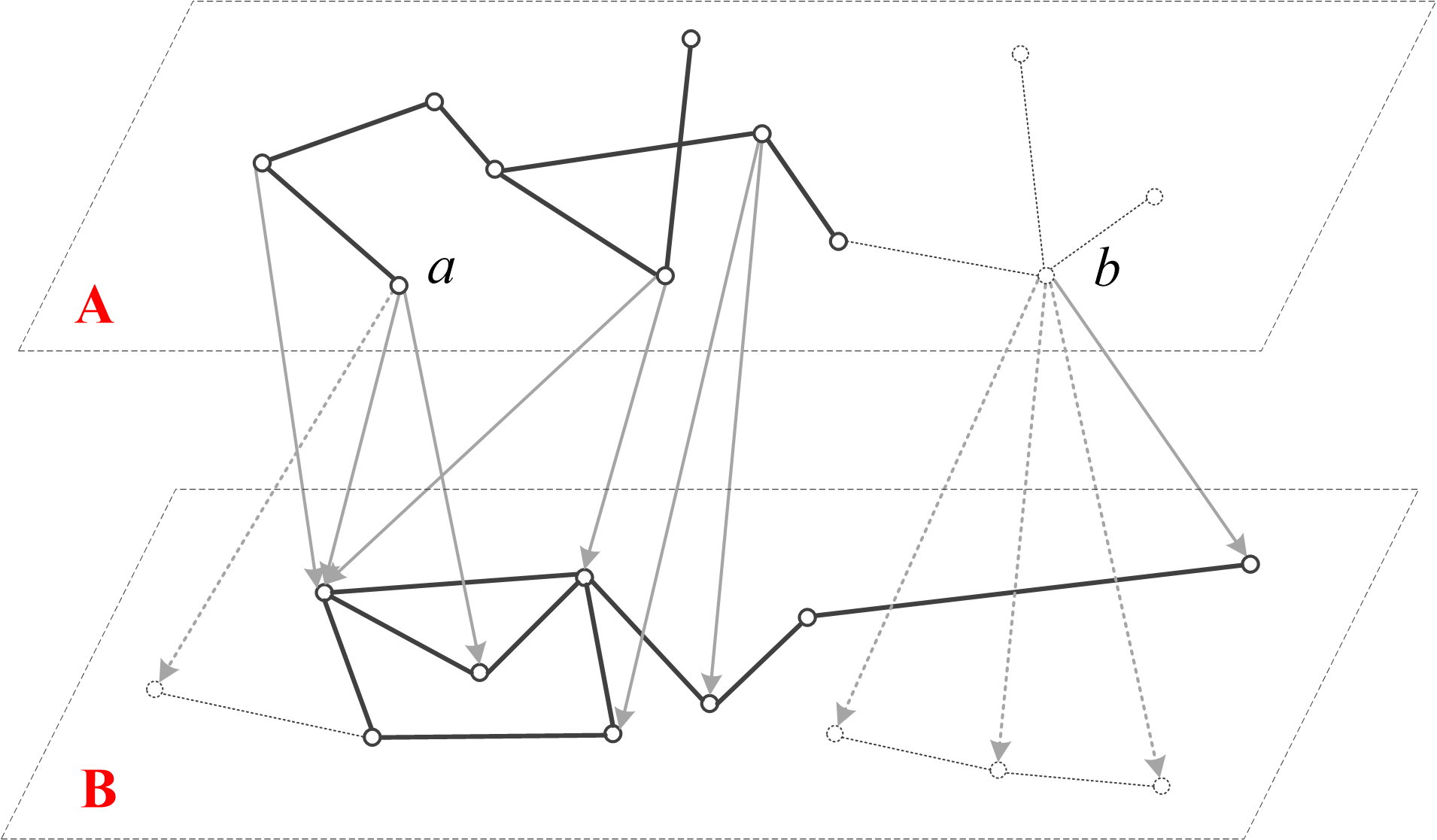}
  \caption{Conditional group dependency failure.}
  \label{fig:multi_yilai}
\end{figure}

There is a default rule in asymmetric dependent failure, that is, the dependent failure will not happen as long as the dependent relation of nodes is maintained, such as node $G_2$ in Figure~\ref{fig:depend}. Based on this, a conditional group-dependent failure model is proposed. This model can tolerate the partial dependent failure of nodes, which is shown in Figure~\ref{fig:multi_yilai}.
As we can see in the figure,  network A is unidirectionally dependent on  Network B, and the scale of the dependency can be different. Let the tolerance limit $\tau$ of the conditional group-dependent failure model be 0.5. For the node $a$ in network A, one of the three nodes it relies on has failed, so the failure ratio is $\frac{1}{3}$, which does not exceed the tolerance limit. Node $a$ will stay normal while node $b$ will have a conditional group-dependent failure because it has a failure ratio of $\frac{3}{4}$ and the value exceeds the tolerance limit.

\subsection{Overload failure model}
Apart from the dependent failure, it is necessary to consider other problems caused by the internal operating mechanism of DGCN. For the physical net, the communication node has its inherent capacity limit. When the load exceeds the tolerable range, it will lead to cascading failure. The same situation occurs on the functional net when the service processing is overloaded. Motter and Lai proposed the `M-L' model~\cite{Motter2004,MLmodel2002} to describe this phenomenon. Based on this, Peng et al.~\cite{Peng2015} presented a cascading failure model considering the load and combined it with the dependent failure model. Hao et al.~\cite{Hao2018} proposed a cascading failure model considering the overload state when facing a traffic jam. Inspired by these, we establish a  cascading failure model considering the overload state for the physical net and functional net, respectively, and then integrate it with the dependent failure model. The specific process of overload cascading failure is: Each node has an initial load at the beginning, when a node fails due to an attack or other reasons, the load of this node will be redistributed according to certain rules. The load-obtained node will fail with a certain probability. If the node fails, a new round of load redistribution and node failure will be triggered. For the physical net, the load is redistributed among the nodes in the entire network, while the load redistribution is limited to nodes of the same type as to the functional net.

\subsubsection{Initial load}
The simplest initial load can be defined as the exponential power of the node degree. However, from the perspective of information interaction, a more reasonable definition of the initial load is the function of the information path, that is, the betweenness of the node~\cite{MLmodel2002}. Since betweenness is the global topology information, for large-scale networks, if the structure is not completely known, it is not easy to obtain the initial load. And the complexity of betweenness calculation is very high, which may not suitable for situations with high real-time requirements. From the perspective of `local definition and local allocation', Wang et al.~\cite{WangJW2009} presented a method for calculating initial loads based on local information.  And it is proved that the proposed initial load which uses the function of the product of the node degree and the degree of neighbors is positively correlated with betweenness. According to the above idea, the initial load of the DGCN is defined separately based on the physical net and the functional net.

The initial load of the physical net node is
\begin{equation}
    L_{\rm{W}}^i(0) = {\left( {{k_i}\sum\limits_{j \in {\Gamma _i}} {{k_j}} } \right)^{{\kappa _W}}},i = 1,2, \cdots {N_{\rm{W}}},
\end{equation}
where $\kappa_{\rm W}$ is the adjustment parameter, which is used to control the initial load distribution of communication nodes; $\Gamma_i$ is the subscript set of neighbors of node $v_i$.

The initial processing load of the functional net node is
\begin{equation}
    L_{\rm{G}}^i(0) = {\left( {{k_i}\sum\limits_{j \in {\Gamma _i}} {{k_j}} } \right)^{{\kappa _G}}},i = 1,2, \cdots {N_{\rm{G}}},
\end{equation}
where $\kappa_{\rm G}$ is the adjustment parameter, which is used to adjust the initial load distribution of functional nodes; $\Gamma_i$ is the subscript set of neighbors of node $v_i$.

\subsubsection{Node capacity}
Due to the cost constraints, there is an upper limit of node capacity in a load-induced network. The usual way to define the node capacity is to assume that it is proportional to its initial load. However, in most real networks, a node with a smaller capacity usually has a larger remaining capacity. The relation between node capacity and load is more likely to be a nonlinear model~\cite{Kim2008,HanHY2017}. In the DGCN of CSOS, if the initial load of the node is large, it indicates that the node is essential. The more important the node is, the more frequently it interacts with other nodes in information or service processing. Therefore, the corresponding residual capacity of the node is small. On the contrary, the load of the less important node is smaller, and there will be more free capacity~\cite{YangYH2017}. Here we adopt the nonlinear model to define the node capacity based on Kim's work \cite{Kim2008}.

The node capacity of the physical net is
\begin{equation}
    C_{\rm{W}}^i = L_{\rm{W}}^i(0) + {\lambda _{\rm{W}}} \cdot L_{\rm{W}}^i{(0)^{{\gamma _{\rm{W}}}}},i = 1,2, \cdots {N_{\rm{W}}},
\end{equation}
where $\lambda_{\rm W}$ and $\gamma_{\rm W}$ are the adjustment parameters.

The node capacity of the functional net is
\begin{equation}
    C_{\rm{G}}^i = L_{\rm{G}}^i(0) + {\lambda _{\rm{G}}} \cdot L_{\rm{G}}^i{(0)^{{\gamma _{\rm{G}}}}},i = 1,2, \cdots {N_{\rm{G}}},
\end{equation}
where $\lambda_{\rm G}$ and $\gamma_{\rm G}$ are the adjustment parameters.

\begin{figure}[!htbp]
  \centering
  \includegraphics[width=0.95\linewidth]{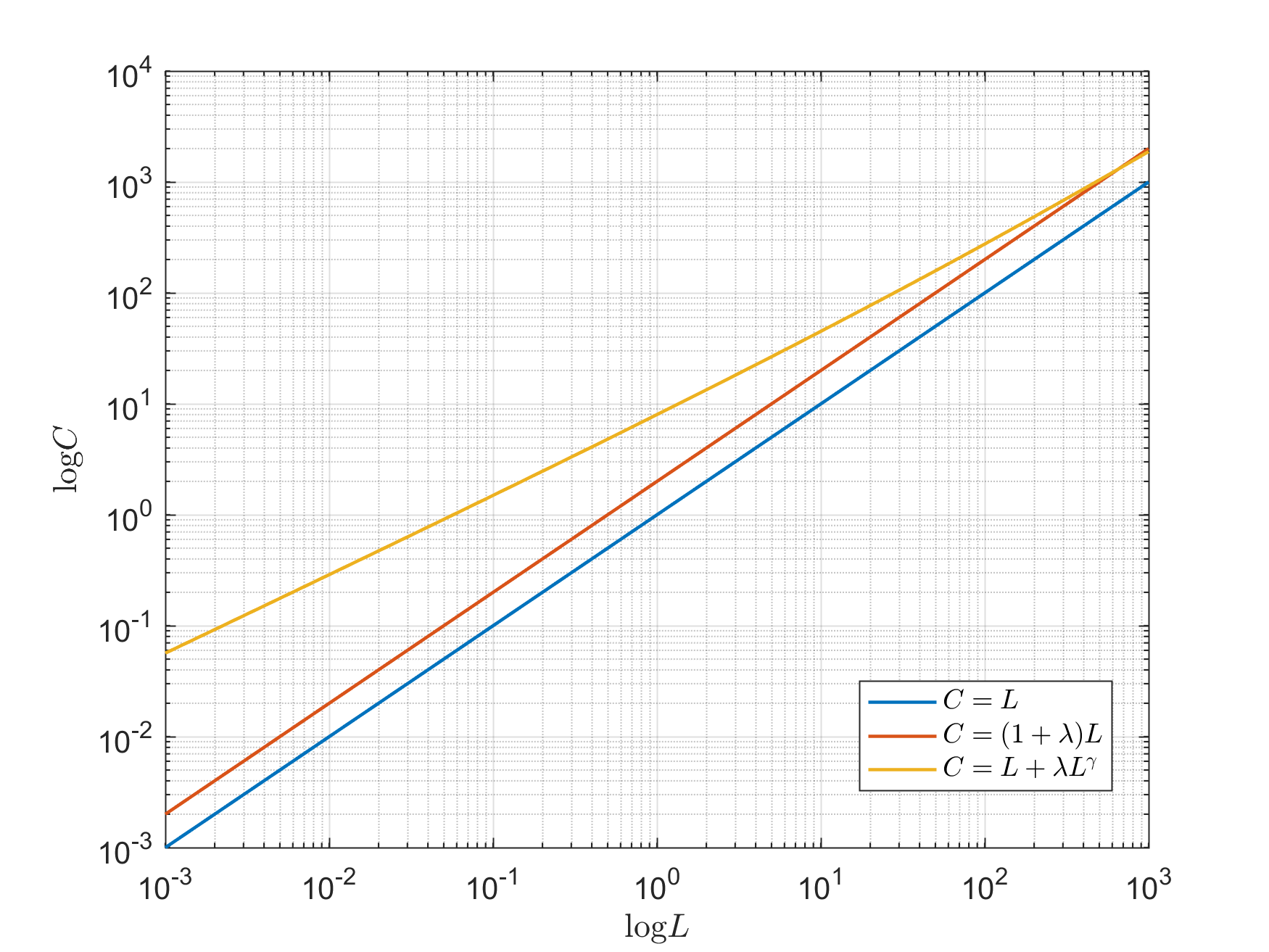}
  \caption{Relation between initial load and capacity.}
  \label{fig:C_L}
\end{figure}

Figure \ref{fig:C_L} presents the relation between the initial load and capacity in nonlinear and linear forms. It can be seen from the figure that the nonlinear model in this paper conforms to the actual situation analyzed above. Especially, the nonlinear model degenerates into a linear model when $\gamma =1$, which shows that it is more general.

\subsubsection{Redistribution strategy of load}
After a node fails, the current load will propagate in a certain way. The common redistribution methods such as local redistribution~\cite{Hong2016}, global redistribution~\cite{Sun2008}, and average distribution~\cite{Moreno2003} have different effects on the robustness of the network when facing cascading failures. For the physical net, the communication transmissions are generally addressed according to the principle of the minimum number of hops,
so local distribution is more inclined to be adopted. In this paper, we proposed a load redistribution strategy from the perspective of the initial static load and subsequent dynamic remaining capacity, which is as follows:

The static allocation ratio for neighbor node $v_j$ of failed node $v_i$ at time 0 is
\begin{equation}
    \Pi _{ij}^s = \frac{{L_{\rm{W}}^j(0)}}{{\sum\limits_{j \in {{\Gamma}_i}} {L_{\rm{W}}^j(0)} }}.
\end{equation}
As for the dynamic allocation at time $t$, we use the ratio based on the dynamic remaining capacity  and it is represented as
\begin{equation}
    \Pi _{ij}^d = \frac{C_{\rm{W}}^j-{L_{\rm{W}}^j(t)}}{{\sum\limits_{j \in {{\Gamma}_i}} C_{\rm{W}}^j-{L_{\rm{W}}^j(t)} }}.
\end{equation}
Therefore, the synthetic allocation ratio is as follows:
\begin{equation}
    \Pi _{\rm{W}}^{ij} = \eta \Pi _{ij}^s + (1 - \eta )\Pi _{ij}^d,
\end{equation}
where $\eta$ is the proportion parameter of two redistribution methods. Then the new obtained load of neighbor node $v_j$ is expressed as
\begin{equation}
    \Delta L_{\rm{W}}^{ij}(t) = \Pi _{\rm{W}}^{ij}L_{\rm{W}}^{i}(t).
\end{equation}
 And we can update the load of the physical net's nodes by the following expression:
\begin{equation}
    L_{\rm{W}}^k(t + 1) = L_{\rm{W}}^k(t) + \sum\limits_{j \in {\Gamma _k}} {\Delta L_{\rm{W}}^{jk}(t)} .
\end{equation}

In the same way, the processing load distribution method of the functional net is easy to obtain. However, it should be noted that the redistribution only occurs among the same type of functional nodes. The node load after a new round of failure is shown as follows:
\begin{equation}
    L_{\rm{G}}^k(t + 1) = L_{\rm{G}}^k(t) + \sum\limits_{j \in {\Gamma _k}} {\Delta L_{\rm{G}}^{jk}(t)} .
\end{equation}

\subsubsection{Failure status judgment}
In the classical `M-L' model~\cite{MLmodel2002}, a node has only two states, namely, the normal state and the failed state. Hao et al.~\cite{Hao2018} proposed a cascading failure model of complex networks considering overloaded nodes. When the load exceeds the capacity of a node, the node will fail with a certain probability in the certain bearing range $\delta$. This state is called the critical state. When the load exceeds the above range, the node fails directly. Motivated by this, the specific rules for judging the failure situation after the node obtains the reallocated load are as follows:

(1) if ${L_i}(t) \le {C_i}$, the node $v_i$ does not fail.

(2) if ${C_i} < {L_i}(t) \le (1 + \delta ){C_i}$, the node $v_i$ is in the critical failure state. Although the load exceeds the capacity of a node, it is still within the affordable range. The more the node is overloaded, the greater the probability of node failure and the failure probability is
    \begin{equation}
        {p_i}(t) = \frac{{{L_i}(t) - {C_i}}}{\delta C_i}.
    \end{equation}
When ${p_i}(t)$ is greater than a random number, the node $v_i$ fails.

(3) if ${L_i}(t) > (1 + \delta ){C_i}$, the node $v_i$ fails immediately.

In summary, taking the cascading failure of the physical net as an example, the failure probability of node $v_i$ is presented as
\begin{equation}\label{fail_p}
    p_{\rm{W}}^i(t) = \left\{ {\begin{array}{*{20}{c}}
    {0,}&{L_{\rm{W}}^i(t) \le C_{\rm{W}}^i}\\
    {\frac{{L_{\rm{W}}^i(t) - C_{\rm{W}}^i}}{\delta C_{\rm{W}}^i},}&{C_{\rm{W}}^i < L_{\rm{W}}^i(t) \le (1 + \delta )C_{\rm{W}}^i}\\
    {1,}&{L_{\rm{W}}^i(t) > (1 + \delta )C_{\rm{W}}^i.}
    \end{array}} \right.
\end{equation}

For the cascading failure of the functional net, the processing load can only spread among nodes of the same type. Once a functional node fails, the processing load of it will redistribute among neighbor nodes according to the same information transmission or service transactions. If the processing load of the current node has been updated, the probability of whether this node fails is similar to Equation~(\ref{fail_p}), and will not be repeated.

\subsection{Attack mode}
In combat confrontation, attacks on combat networks are generally divided into random attacks and intended attacks~\cite{GaoYL2018}. The random attack is to randomly select several nodes to make them invalid, while the intended attack is to make nodes invalid according to a certain order of node importance. Since DGCN is a double-layer network, there are different attack modes for different networks. For example, in different combat phases, the enemy may select specific functional net nodes to attack or destroy the nodes of the physical net randomly. According to the attack object of the combat network and the attack intention, the attack mode can be divided into the following six types:

(1) Random Single Physical Network Attack~(RSPA), randomly select nodes with a ratio $f$ from the physical net to attack and let these nodes fail.

(2) Intended Single Physical Network Attack~(ISPA), select nodes with a ratio $f$ from the physical net according to the descending node degree, and let them fail by attacking.

(3) Random Single Functional Network Attack~(RSFA), randomly select nodes with a ratio $f$ from the physical net to attack and let these nodes fail.

(4) Intended Single Functional Network Attack~(ISFA), select nodes with a ratio $f$ from the functional net according to the descending node degree, and let them fail by attacking.

(5) Random Double Networks Attack~(RDA), randomly select nodes with a ratio $0.5f$, respectively,  from the physical and functional net to attack, and let these nodes fail.

(6) Intended Double Networks Attack~(IDA), select nodes with a ratio $0.5f$, respectively, from the physical net and functional net according to the descending node degree, and let them fail by attacking.

\begin{figure}[!htbp]
  \centering
  \includegraphics[width=0.95\linewidth]{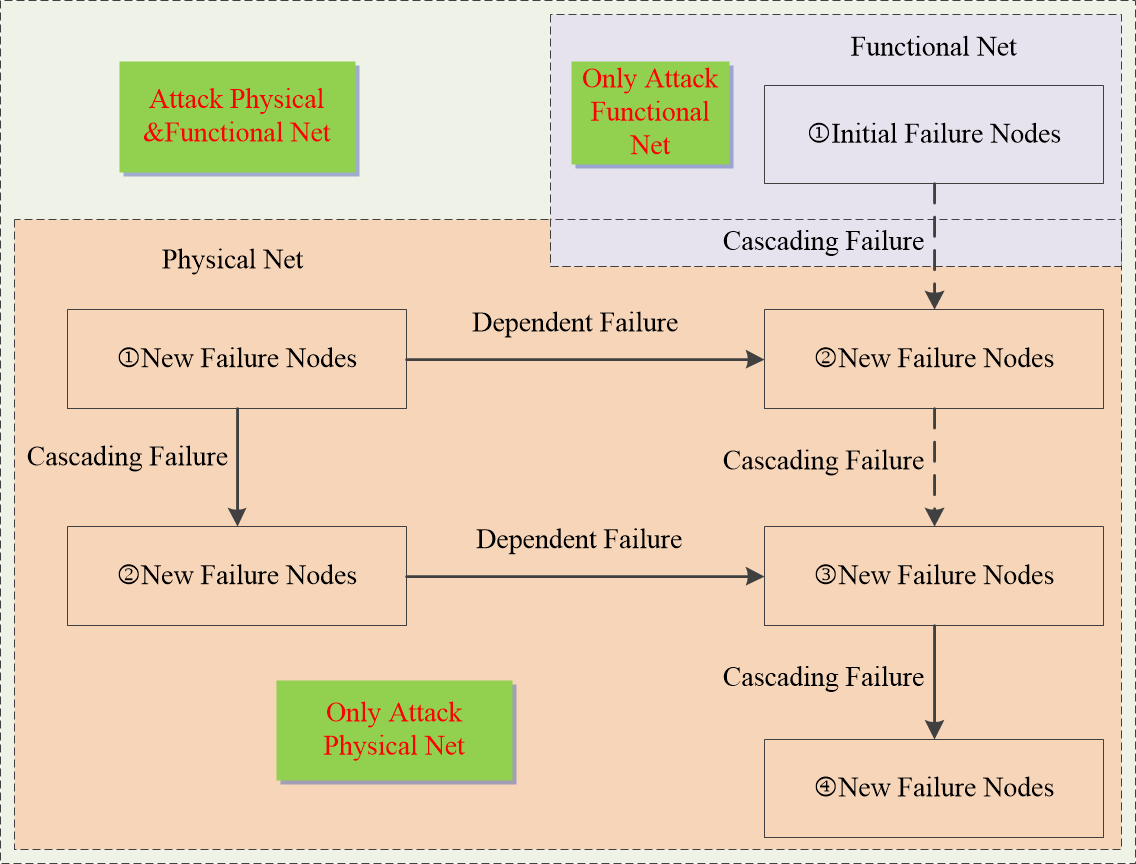}
  \caption{Failure process under different attack modes.}
  \label{fig:failprocess}
\end{figure}

Different attack modes may have different effects on robustness. Combining the asymmetric dependent failure model and overload cascading failure model, the failure process of the DGCN of CSOS can be represented as Figure~\ref{fig:failprocess}, in which the dashed boxes of three different colors represent different attacked objects. These nodes in a double-layer network will suffer random attacks or intended attacks. As can be seen from the figure, the failure process of the physical net always leads to the failure of the functional net, so the functional net is more fragile. Any disturbance in the network may cause serious consequences of `failure, disconnection and paralysis'.

\section{Simulation experiments on robustness}
\label{sec:4}

\subsection{Robustness evaluation index}
The general definition of the robustness of the combat network is the ability of a system to maintain its original functions, characteristics, and organizational structure under external disturbances~\cite{DiP2011}.
In this paper, the robustness of DGCN for CSOS refers to the ability of a system to continue maintaining combat performance after some nodes failed in the network. Due to the existence of a functional net in DGCN, the performance evaluation can be more real. Here we define the robustness evaluation index from two aspects:

The first aspect of the robustness measure is based on combat network topology. The scale of the maximal component is selected as an index of the damage effect of the combat network. One individual component is a subgraph with connectivity and isolation in the network. As a result, the maximal component is a subgraph with the largest node scale in the network, which is denoted as $S_{\rm huge}$. With the increase of the scale of the maximal component in the combat network, the interconnections among combat network nodes get closer and the information flow efficiency of the combat network becomes higher. In this paper, we adopt the approach of node shrinking to iteratively calculate the scale of the maximal component~\cite{DengY2016}. The initial maximal component of DGCN is

\begin{equation}
    S_{huge}(G) = {N_{\rm{G}}} + {N_{\rm{W}}},
\end{equation}
where $N_{\rm{G}}$ and $N_{\rm{W}}$ are the node number of the functional net and the physical net, respectively.

\begin{figure}[!htbp]
  \centering
  \includegraphics[width=0.9\linewidth]{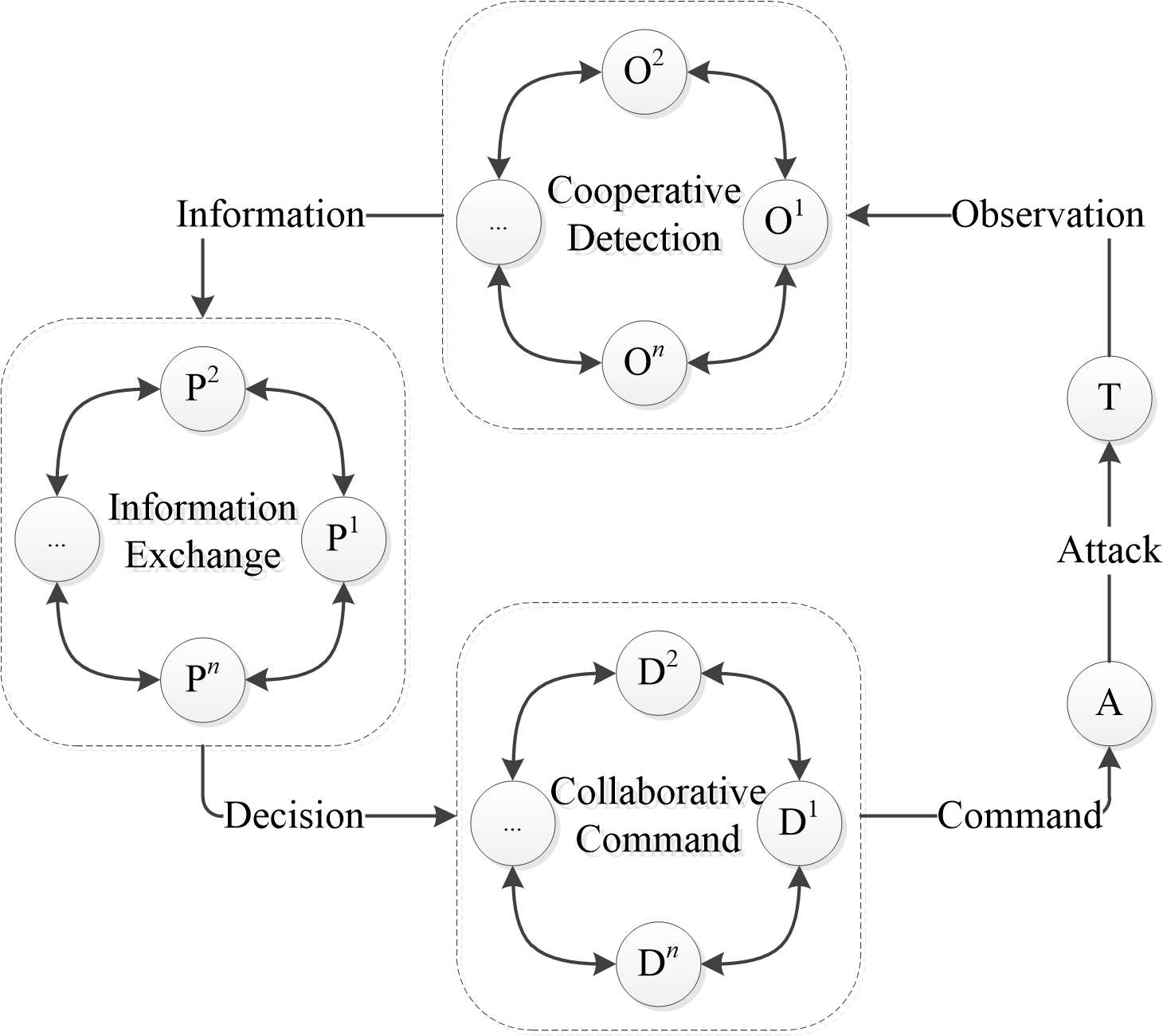}
  \caption {Generalized CELP diagram.}\label{fig_loop}
\end{figure}

The second aspect of the robustness evaluation index is based on the operational capability of the combat network. The number of kill links can be used to describe the operational capability of a combat network~\cite{ZhaoDL2019}. In the case of a DGCN of CSOS, the number of combat effectiveness links~(CELKs) $S_{\rm links}$ is introduced to measure the operational capability. According to the idea of Boyd's OODA cycle~\cite{Boyd2018}, the combat network exerts operational capability by forming a CELK of  `intelligence obtaining--intelligence processing--commanding and decision--attack and damage' around the target,
namely, the OODA kill link. For a more general CELK, the mutual coordination among the intelligence obtaining node O, the intelligence processing node P, and the commanding and decision node D should also be taken into consideration. The flow of generalized CELK with a target can be demonstrated as the generalized combat efficiency loop~(CELP) (see  Figure~\ref{fig_loop}).

 Since the nodes cooperation in generalized CELK may lead to an infinite long link, seven types of CELKs commonly used in practice are selected as the basis for quantity calculation~\cite{YJT2022}. The detailed description of those CELKs is indicated in Table~\ref{tab_links}.

\begin{table*}[!htbp]
\caption{Seven types of common CELKs and their definitions. \label{tab_links}}
\begin{tabularx}{\textwidth}{p{4cm}p{8cm}}
  \toprule
  \centering \textbf{CELK}	    &  \textbf{Definition}	\\
  \midrule
  \centering O--C--P--C--D--C--A	    	& The standard CELK		\\
  \centering O--C--O--C--P--C--D--C--A   	& CELK with cooperative detection	\\
  \centering O--C--P--C--P--C--D--C--A	    & CELK with information interaction	\\
  \centering O--C--P--C--D--C--D--C--A		& CELK with coordinated command		\\
  \centering O--C--O--C--P--C--P--C--D--C--A	& CELK with cooperative detection and information interaction	\\
  \centering O--C--O--C--P--C--D--C--D--C--A	& CELK with cooperative detection and coordinated command		\\
  \centering O--C--O--C--P--C--P--C--D--C--D--C--A	& CELK with cooperative detection, coordinated command and information interaction			\\
  \bottomrule
\end{tabularx}
\end{table*}

The accessibility matrix $\tilde{\bm{S}}$ of the entire combat network is calculated according to the adjacency matrix $\bm{S}$ of the functional net:
\begin{equation}\label{keda_matrix}
 \begin{array}{l}
   {(\bm{S} + \bm{I})^{(1)}} \ne {(\bm{S} + \bm{I})^{(2)}} \ne  \cdots \\
   {\qquad\qquad}\ne {(\bm{S} + \bm{I})^{(r)}} = {(\bm{S} + \bm{I})^{(r + 1)}} = \tilde{\bm{S}}\,,
 \end{array}
\end{equation}
where $I$ is the identity matrix. Equation~(\ref{keda_matrix}) is the power boolean operation on $(\bm{S}+\bm{I})$, and $r+1$ is the times of power multiplication. The connectivity between intelligence obtaining and attack/damage nodes can be known based on the accessibility matrix. For any node $O_{\rm i}$ and node $A_{\rm j}$, if $\tilde {\bm{S}}(i,j)=1$, then $O_{\rm i}$ can reach $A_{\rm j}$ according to a path with practical meaning. Assuming $\bm{S}(j,i)=1$, then the number of CELPs is the trace of the product of the corresponding nodes' accessibility matrices, which is also the number of CELKs. Because of the dependency of functional net on the physical net, the information transmission must go through the communication node as shown in Figure~\ref{fig:flow}. The CELK should also correspondingly consider the role of communication nodes. Taking the `O$\to $P$\to $D$\to $A' link as an example, the number of this link can be calculated by
\begin{equation}
    \begin{array}{l}
    S_{{\rm{OPDA}}}^{{\rm{WG}}}(G) = {\rm{tr}}\{ [{\bm{S}_{{\rm{OP}}}} \wedge ({\bm{S}_{{\rm{OC}}}} \times {\bm{S}_{{\rm{CC}}}} \times {\bm{S}_{{\rm{CP}}}})] \times \\ 
    {\qquad\qquad\qquad }[{\bm{S}_{{\rm{PD}}}} \wedge ({\bm{S}_{{\rm{PC}}}} \times {\bm{S}_{{\rm{CC}}}} \times {\bm{S}_{{\rm{CD}}}})] \times \\
    {\qquad\qquad\qquad }[{\bm{S}_{{\rm{DA}}}} \wedge ({\bm{S}_{{\rm{DC}}}} \times {\bm{S}_{{\rm{CC}}}} \times {\bm{S}_{{\rm{CA}}}})] \times {\bm{S}_{{\rm{AO}}}}\} ,
    \end{array}
\end{equation}
Where $\wedge$ is the boolean operation of two matrices, that is, if the corresponding element $(i, j)$ in matrix $\bm{X}$ and $\bm{Y}$ are both greater than 0, then  $\bm{X}(i,j) \wedge \bm{Y}(i,j) = 1$. Then the total number of links is calculated by
\begin{equation}
    S_{\rm links}(G) = \sum\limits_{i = 1}^7 {S_{{\rm link}_i}}(G) .
\end{equation}

The robustness of the combat network in this paper is calculated by relative metrics. For the original combat network $G$ which has not been attacked, the initial largest component scale is $S_{\rm huge}(G)$. The amount of CELKs is $S_{\rm links}(G)$. For the attacked combat network $G'$, the corresponding largest component scale and the number of CELKs are $S_{\rm huge}(G')$ and $S_{\rm links}(G')$, respectively. We can measure the robustness of the combat network in the following expression:

\begin{equation}\label{eq_R}
    R = {\left( {\frac{{S_{\rm links}}(G')}{{S_{\rm links}(G)}}} \right)^\alpha }{\left( {\frac{{S_{\rm huge}}(G')}{{S_{\rm huge}(G)}}} \right)^{1 - \alpha }},
\end{equation}
where $\alpha$ is a proportion parameter indicating the preference for two different metrics. The default value of $\alpha$ is 0.5.

\subsection{Experiment results and analysis}
In order to study the robustness of the DGCN of CSOS, the simulation experiment was carried out by using the model network. Model networks such as ER random network~\cite{Erdos1984}, Goh scale-free network with tunable parameter~\cite{Goh2001}, and NW small world network\cite{Newman1999} are selected as functional net and physical net, respectively. The network scale of functional net is $N_{\rm G}=150$, where $N_{\rm O}=50$, $N_{\rm P}=40$, $N_{\rm D}=30$, $N_{\rm A}=30$, and the scale of physical net is $N_{\rm W}=100$. As for the parameter setting of the model network, the connection probability among different nodes in the ER-net is ${p_{{\rm{OO}}}} = 0.02$, ${p_{{\rm{OP}}}} = 0.03$, ${p_{{\rm{PP}}}} = 0.05$, ${p_{{\rm{PD}}}} = 0.03$, ${p_{{\rm{DD}}}} = 0.05$, ${p_{{\rm{DA}}}} = 0.03$, ${p_{{\rm{AA}}}} = 0.03$, ${p_{{\rm{CC}}}} = 0.07$. The power exponent of Goh-net is $\beta = 2.3$, and the average degree of the network is $\left\langle k \right\rangle = 6$. The functional nodes with different types are connected according to the parameters of the ER-net. The parameters of NW-net are $K=2$, ${p_{{\rm{OO}}}} = 0.08$, ${p_{{\rm{PP}}}} = 0.1$, ${p_{{\rm{DD}}}} = 0.14$, ${p_{{\rm{AA}}}} = 0.14$ and  ${p_{{\rm{CC}}}} = 0.05$. The remaining nodes are also connected according to the parameters of the ER-net. The functional net one-way multiply depends on the physical net, and the scale of group dependency is always 5.
To reduce the randomness in the experiment, each type of the above network is repeatedly generated 300 times according to the given parameters. When conducting simulation experiments, unless otherwise specified, the following default parameters are uniformly used: $\tau  = 0.8$, ${\kappa _{\rm{G}}} = {\kappa _{\rm{W}}} = 0.5$, ${\lambda _{\rm{G}}} = {\lambda _{\rm{W}}} = 1$, ${\gamma _{\rm{G}}} = {\gamma _{\rm{W}}} = 1.1$, ${\delta _{\rm{G}}} = {\delta _{\rm{W}}} = 0.3$. The attack mode is IDA, and the initial failure ratio $f$ ranges from 0 to 0.4.

The simulation software is Matlab 2016b with Windows10, and the hardware configuration is Intel(R) Core(TM) i7-10750H CPU @ 2.60GHz.

\subsubsection{Robustness with different attack modes}
\begin{figure*}[!htbp]
  \centering
  \includegraphics[width=1\textwidth]{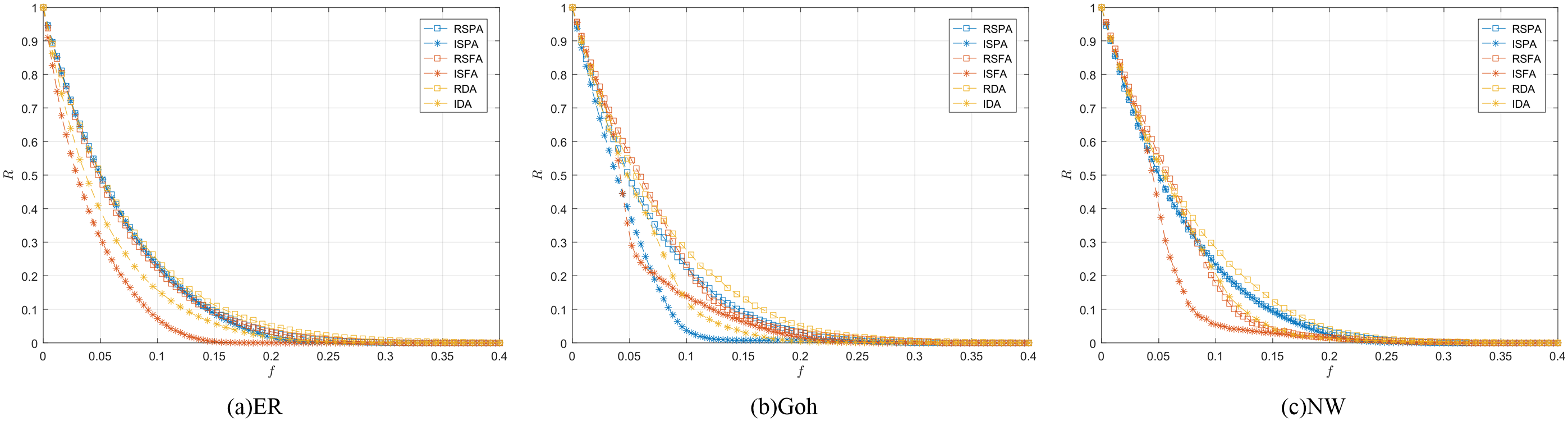}
  \caption{Robustness with different attack modes.}
  \label{fig:attackmode}
\end{figure*}

On the basis of the default parameters, change the attack mode to make the nodes fail in the DGCN of CSOS. The combat networks are attacked when the physical net and the functional net are ER-net, Goh-net and NW-net, respectively. To study the influence of asymmetric attack on the robustness of the DGCN of CSOS, the variations of the robustness of the combat networks under six different attack modes are shown in Figure~\ref{fig:attackmode}.
  As can be seen from the figure, the robustness of different attack modes with different model network structures is quite different.
For attacks on the physical net and when the model network is the ER-net and the NW-net, whether it is a random or intended attack mode, the impact on the robustness of the system is almost the same.
  When the model network is Goh-net, the deliberate attack on the physical net has the greatest impact on the robustness of the DGCN of CSOS although the ISFA mode has a better attack effect for a short period of time.
For attacks on the functional net and when the model network is the ER net and the NW net, the influence on robustness caused by ISFA mode is the greatest.
  When the model network is the Goh-net, the effect of ISFA mode is not as good as that of ISPA mode.
Generally, deliberate attacks on double-layer networks have an impact on the robustness between that with ISPA mode and ISFA mode. For several types of network models to be attacked in different modes, the influence on the robustness of the combat network is complex, but the common feature is that the impact of deliberate attacks on the robustness is greater than that of random attacks, and random attacks on the double-layer network have a negative impact on the combat network's robustness. Therefore, it is necessary to fully understand the architecture of the attacked object during combat, so that the attack strategy can be reasonably developed to achieve a better combat effect.

\subsubsection{Robustness with different tolerance parameter $\tau$}
\begin{figure*}[!htbp]
  \centering
  \includegraphics[width=1\textwidth]{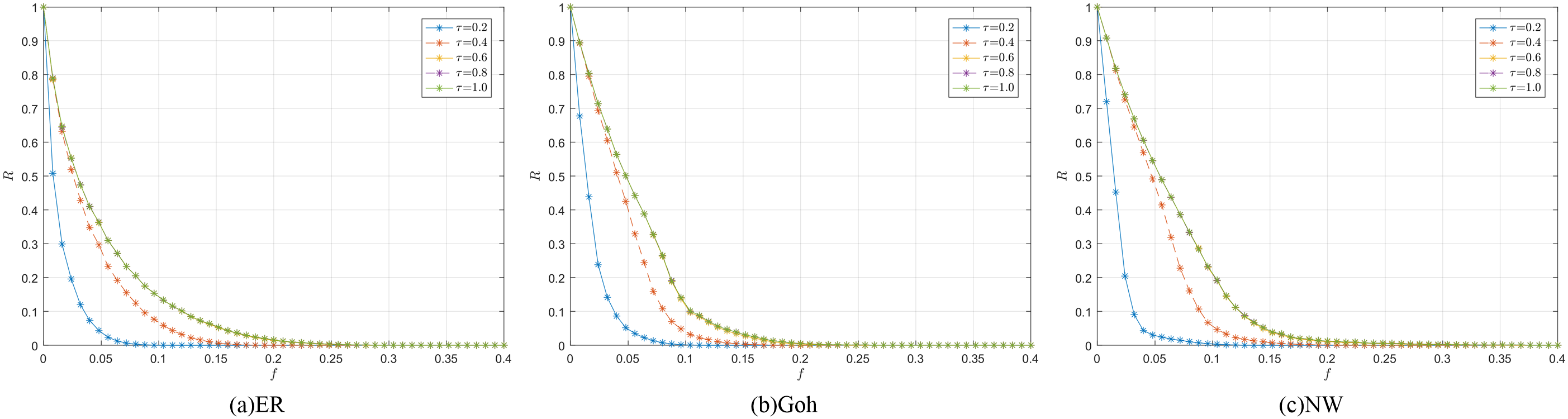}
  \caption{Robustness with tolerance parameter $\tau$.}
  \label{fig:tau}
\end{figure*}

In order to explore the influence of different tolerances on the dependence characteristics of the DGCN of CSOS, the tolerance coefficient is varied in this experiment. Because the dependency scale of the group dependency is always 5, the value of the tolerance coefficient is varying at an interval of 0.2. And the robustness of the DGCN of CSOS when different model networks are used for simulation is shown in Figure~\ref{fig:tau}. It can be seen from the figure that the experiments under the three model networks have a consistent conclusion: the smaller the tolerance coefficient, the worse the robustness of the system. When the tolerance coefficient reaches a certain threshold, there is a certain upper limit for the change of robustness and the robustness drop curves no longer improve.

\subsubsection{Robustness to different parameter $\kappa$}
\label {subsec:kappa}

\begin{figure*}[!htbp]
  \centering
  \includegraphics[width=1\textwidth]{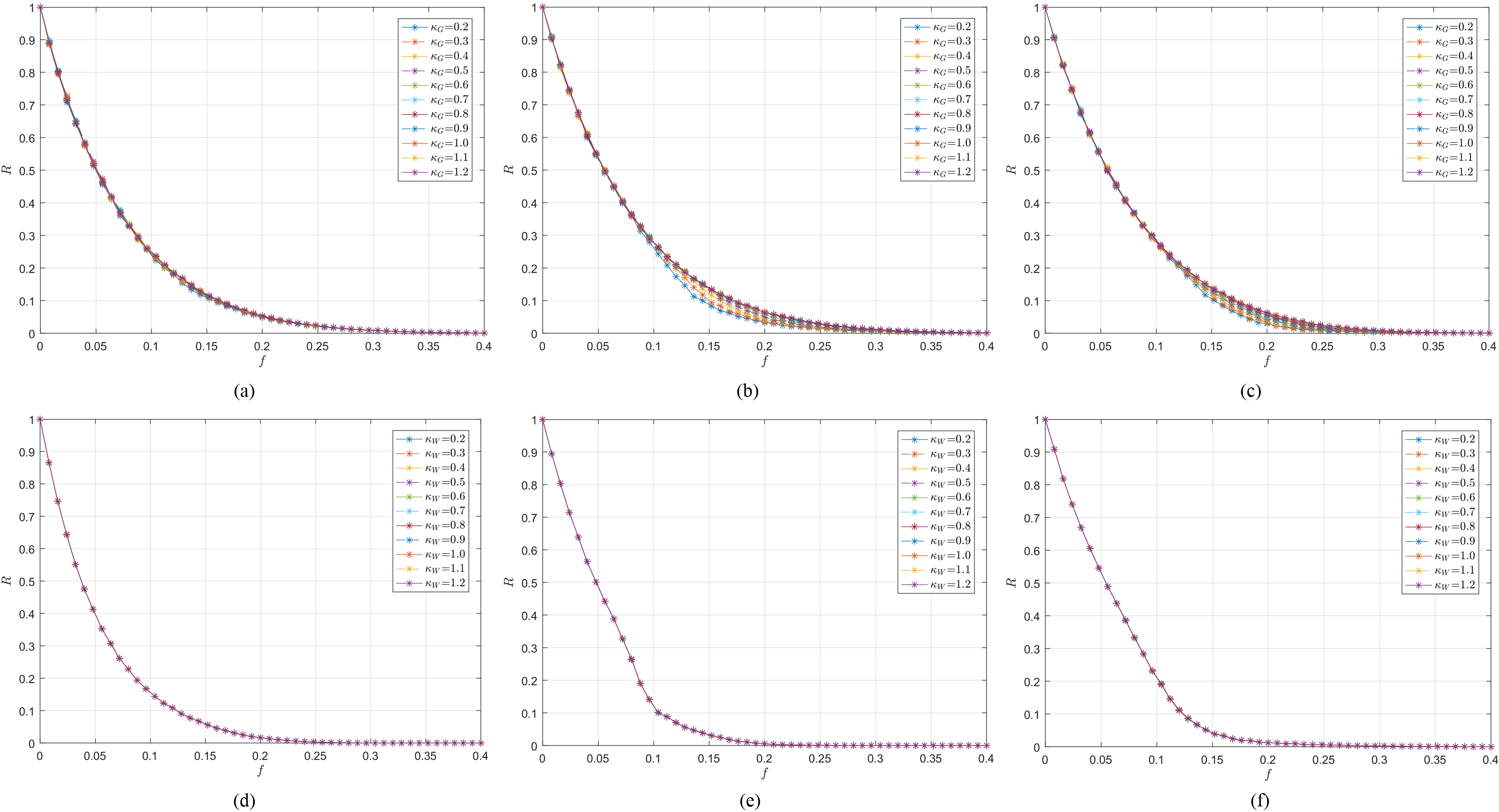}
  \caption{Robustness with different $\kappa$. (a), (b), (c) represent the robustness under different $\kappa _{\rm G}$ when the subnet is ER-net, Goh-net and NW-net, respectively; (d), (e), (f) represent the robustness under different $\kappa _{\rm W}$ when the subnet is ER-net, Goh-net and NW-net, respectively.}
  \label{fig:kappa}
\end{figure*}

Changing only the initial load parameters and keeping other parameters stable, we discuss the influence of parameter $\kappa$ on the robustness of the DGCN of CSOS. Let the load parameter $\kappa _{\rm G}$ of the functional net vary from 0.2 to 1.2 at an interval of 0.1, and the parameter $\kappa _{\rm W}$ of the physical net also changes according to this rule. Simulation experiments are carried out with different parameters for two different hierarchical networks, and the results are shown in Figure~\ref{fig:kappa}. It can be seen from the figure that for the functional subnet, the robustness of the system improves with the increase of $\kappa _{\rm G}$, but the performance improvement is smaller and smaller. When the initial failure ratio is small, the robustness decline curves almost overlap. As the failure ratio increases, the difference becomes obvious. This is because the parameter $\gamma$ of the node capacity is greater than 1, and the node capacity increases faster as the initial load increases, which appends the remaining capacity. Therefore, the overload cascading failure with a small initial failure ratio is alleviated to a certain extent.
  However, the robustness curve distinction is not obvious for the combat network with ER-net as the model network. Because the random connection makes the degree distribution more uniform, the node load and capacity are correspondingly more uniform and the difference of attack effect is not obvious.
As for the physical net, the robustness curves of the DGCN of CSOS almost completely coincide with the increase of $\kappa _{\rm W}$. This is because the physical net plays the role of intermediary transmission. We have found that when the R reaches 0 for the first time, the maximal component of the physical net is still larger than the half of initial scale. For these homogeneous nodes of the physical net, as long as the nodes depended on the functional net are still in the maximal component of the physical net, the robustness of the combat network will not change. On the contrary, node failure of the functional net will have a great impact on the robustness.

\subsubsection{Robustness to different parameter $\lambda$}
\label{subsec:lambda}

\begin{figure*}[!htbp]
  \centering
  \includegraphics[width=1\textwidth]{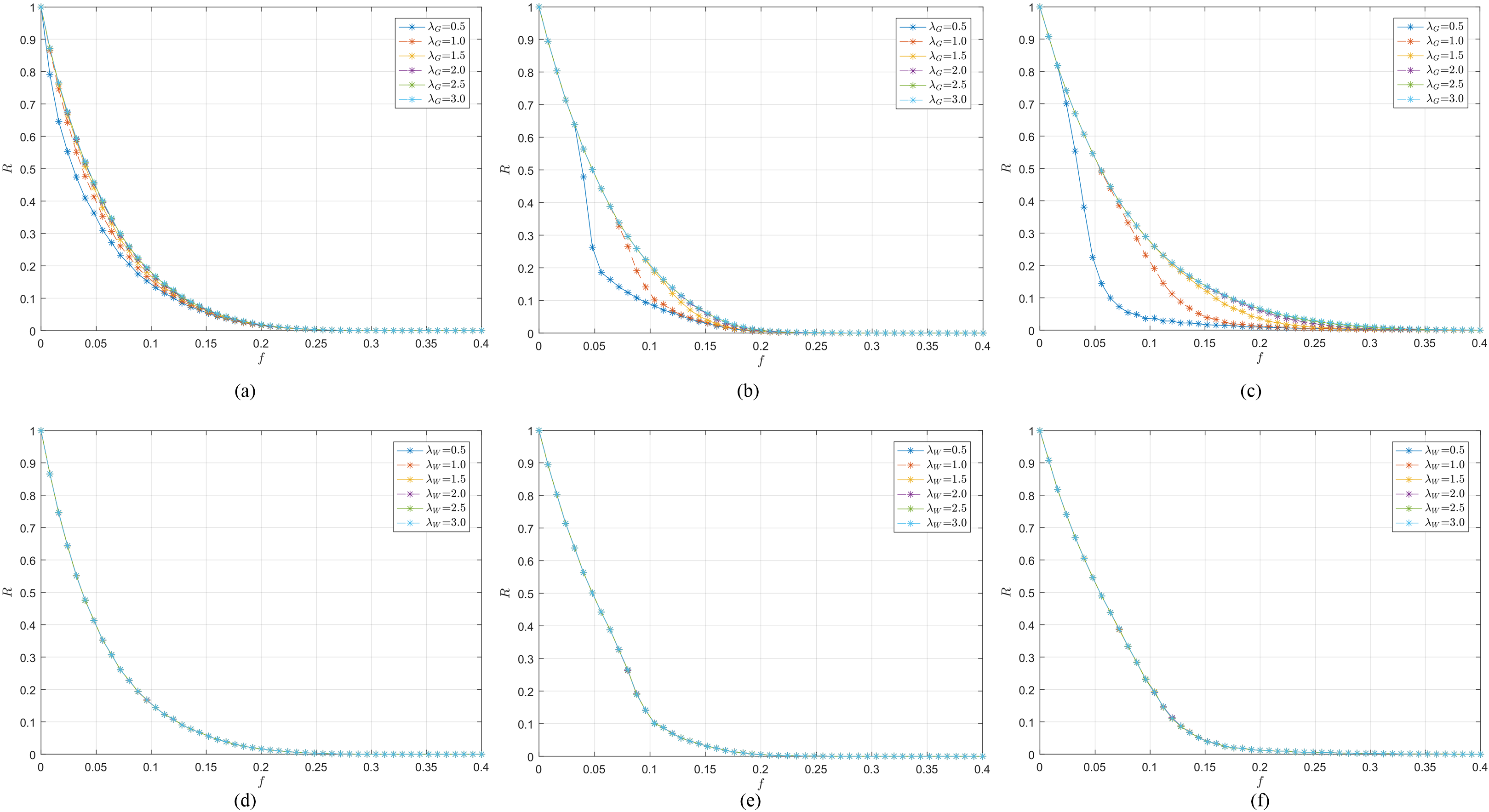}
  \caption{Robustness with different $\lambda$. (a), (b), (c) represent the robustness under different $\lambda _{\rm G}$ when the subnet is ER-net, Goh-net and NW-net, respectively; (d), (e), (f) represent the robustness under different $\lambda _{\rm W}$ when the subnet is ER-net, Goh-net and NW-net, respectively.}
  \label{fig:lambda}
\end{figure*}

Let the other parameters be fixed apart from the linear parameter of the node capacity. We examine the influence of parameter $\lambda$ on the robustness of the DGCN of CSOS. Let the parameter $\lambda _{\rm G}$ of the functional net change in [0.5, 3] at an interval of 0.5. Similarly, the parameter $\lambda _{\rm W}$ of the physical net also varies according to this rule. Simulation experiments are carried out with different parameters for two different hierarchical networks, and the results are shown in Figure~\ref{fig:lambda}. It can be seen from the figure that for the functional net, the robustness of the system improves with the increase of parameter $\lambda _{\rm G}$, indicating that increasing the node capacity of the functional net can strengthen the robustness of the system. The magnitude of improvement for different model networks is NW-net $>$ Goh-Net $>$ ER-Net. When the initial failure proportion is small, the robustness-decreasing curves are almost the same, and the proportion of overlap increases with the accumulation of $\lambda _{\rm G}$, indicating that there is an upper limit on the robustness-decreasing curve. As the failure proportion increases, the difference under different conditions is gradually obvious, and then gradually approaches 0 due to the reduction in robustness. For the physical net, the robustness curves of the system almost completely coincide with the increase of $\lambda _{\rm W}$, the reason is the same as that in Section~\ref{subsec:kappa}.

\subsubsection{Robustness to different parameter $\gamma$}
\label{subsec:gamma}

\begin{figure*}[!htbp]
  \centering
  \includegraphics[width=1\textwidth]{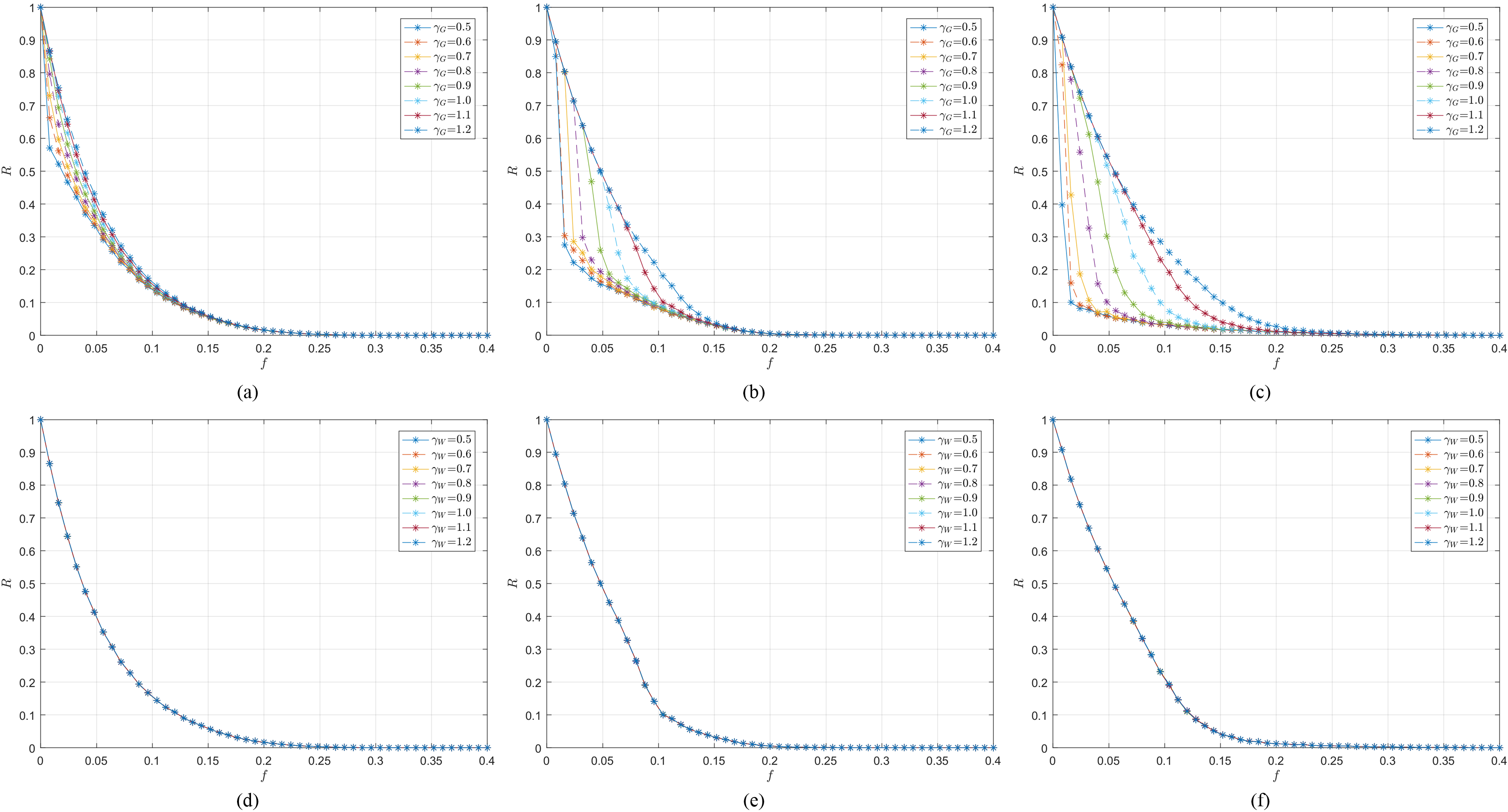}
  \caption{Robustness with different $\gamma$. (a), (b), (c) represent the robustness under different $\gamma _{\rm G}$ when the subnet is ER-net, Goh-net and NW-net, respectively; (d), (e), (f) represent the robustness under different $\gamma _{\rm W}$ when the subnet is ER-net, Goh-net and NW-net, respectively.}
  \label{fig:gamma}
\end{figure*}

When other parameters remain unchanged, only the nonlinear parameter $\gamma$ of node capacity is changed, and the influence of this parameter on the robustness of the DGCN of CSOS is investigated. Let the load parameter $\gamma _{\rm G}$ of the functional net change in [0.5, 1.2] at an interval of 0.1. Similarly, the parameter $\gamma _{\rm W}$ of the physical net also varies according to this rule. Simulation experiments are carried out with different parameters for two different hierarchical networks, and the results are shown in Figure~\ref{fig:gamma}. The relevant laws can be analyzed from the figure, and the conclusions are similar to those in Section~\ref{subsec:lambda}.

\subsubsection{Robustness to different parameter $\delta$}

\begin{figure*}[!htbp]
  \centering
  \includegraphics[width=1\textwidth]{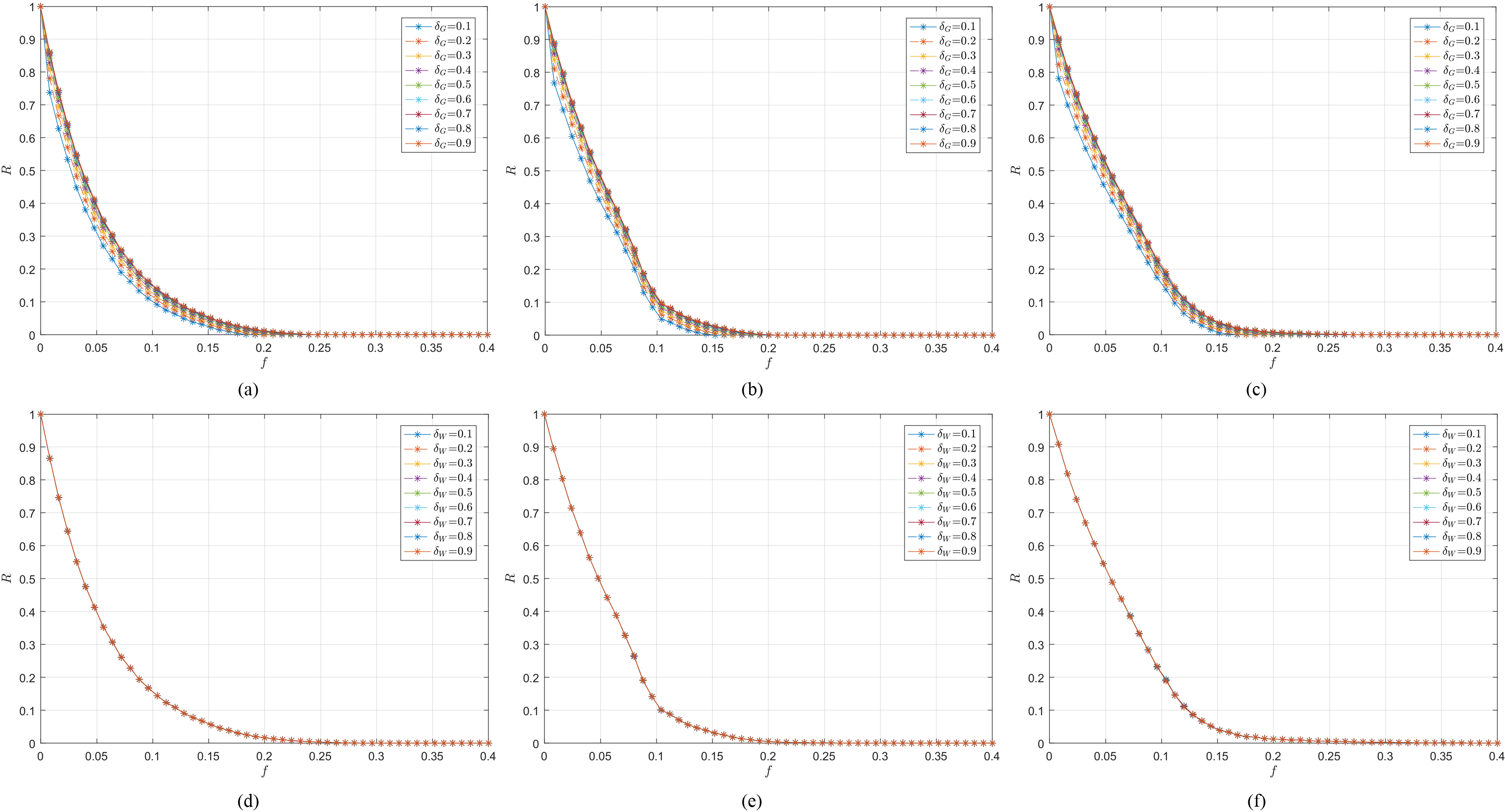}
  \caption{Robustness with different $\delta$. (a), (b), (c) represent the robustness under different $\delta _{\rm G}$ when the subnet is ER-net, Goh-net and NW-net, respectively; (d), (e), (f) represent the robustness under different $\delta _{\rm W}$ when the subnet is ER-net, Goh-net and NW-net, respectively.}
  \label{fig:delta}
\end{figure*}

Apart from the attack mode, tolerance limit of dependent failure and parameters of load and capacity, the robustness of the DGCN of CSOS is also affected by the endurance parameter $\delta$. Let the endurance parameter $\delta _{\rm G}$ of the functional net change in [0.1, 0.9] at an interval of 0.1, and the parameter $\delta _{\rm W}$ of the physical net also change according to this rule. Simulation experiments are carried out with different parameters for two different hierarchical networks, and the results are shown in Figure~\ref{fig:delta}. As can be seen from the figure, for the functional net, the smaller the $\delta _{\rm G}$ is, the less robust the combat network is. With the increase of $\delta _{\rm G}$, that is, the endurance proportion of the overload state increases, the probability of failure in the overload state will decrease. So the robust performance of the system will correspondingly improve. For the physical net, the robustness curves of the system almost completely coincide with the increase of $\delta _{\rm W}$, the reason is also the same as that in Section~\ref{subsec:kappa}.

\section{Conclusions}
\label{sec:5}
The phenomenon of cascading failure in the DGCN of CSOS will lead to a complete collapse of the system in the face of a relatively small proportion of damage, resulting in the loss of combat effectiveness. Finding out the reasons that affect the robustness of the combat network to improve the system capability and further stop the collapse is of great significance. In this paper, we have contributed some results as follows:

(1) We have established a more realistic combat network model of the DGCN of CSOS;

(2) We have designed the cascading failure model of the DGCN by combining the asymmetric dependent failure, conditional group-dependent failure and overload failure together. We have also designed the load reallocation strategy and a more practical robustness index;

(3)~We have investigated the robustness of the combat network with six different attack modes and different model parameters. The simulation results show that the robustness of the combat network can be effectively improved by improving the tolerance limit of one-way dependency of the functional net, the node capacity of the functional net and the tolerance of the overload state. When the attack intensity remains steady, the combat network's ability to deal with deliberate attacks is weaker than that with random attacks. And for different attack methods, different model networks perform inconsistently. It is necessary to design a reasonable network structure to enhance the anti-destruction ability of the combat network.

The model established in this paper is more real, and the laws concluded from the simulation experiments have certain reference significance for optimizing the structure of the combat network and improving the robustness of CSOS.
However, there are still some limitations to our study. For one thing, combat networks are modelled statically and are not considered for modelling networks in dynamic situations; for another, there is a lack of data on real combat networks for reasons of secrecy. Therefore, more practical and complex works need to be done in future research, such as the dynamic model of the combat network, the effectiveness of the model on real data, 
the impact of collocation of different model networks on robustness, and the influence of multiple factors on system robustness.

\bibliographystyle{ws-jcsc}   
\bibliography{sample}  








\end{document}